\documentclass[sigconf]{acmart}

\usepackage{graphicx}
\usepackage{subcaption}
\usepackage{hyperref}
\usepackage{xcolor}
\usepackage{xspace}

\AtBeginDocument{%
  }

\copyrightyear{2026}
\acmYear{2026}
\setcopyright{cc}
\setcctype{by}
\acmConference[ICSE-SEIP '26]{2026 IEEE/ACM 48th International Conference on Software Engineering}{April 12--18, 2026}{Rio de Janeiro, Brazil}
\acmBooktitle{2026 IEEE/ACM 48th International Conference on Software Engineering (ICSE-SEIP '26), April 12--18, 2026, Rio de Janeiro, Brazil}
\acmPrice{}
\acmDOI{10.1145/3786583.3786876}
\acmISBN{979-8-4007-2426-8/2026/04}

\begin{document}

\newcommand{\ing}{\ifdefined\DOUBLEBLIND ABC\else ING\fi\xspace}


\definecolor{ingorange}{RGB}{190,230,255}

\newcounter{findingcount}

\newcommand{\finding}[1]{%
  \refstepcounter{findingcount}%
  \par\vspace{2pt}\noindent%
  \begingroup%
    \setlength{\fboxrule}{0.6pt}
    \setlength{\fboxsep}{4pt}
    \fcolorbox{ingorange!75!black}{ingorange!5!white}{%
      \parbox{\dimexpr\linewidth-2\fboxsep-2\fboxrule\relax}{%
        \small\textbf{Finding~\thefindingcount.}\ \emph{#1}%
      }%
    }%
  \par\endgroup%
}

\title[Learning from Change: Predictive Models for Incident Prevention in a Regulated IT Environment]{Learning from Change: Predictive Models for Incident Prevention\\in a Regulated IT Environment}

\author{Eileen Kapel}
\email{Eileen.Kapel@ing.com}
\orcid{0000-0002-7136-398X}
\affiliation{%
  \institution{ING Bank}
  \city{Amsterdam}
  \country{The Netherlands}
}

\author{Jan Lennartz}
\affiliation{%
  \institution{ING Bank}
  \city{Brussels}
  \country{Belgium}}
\email{Jan.Lennartz@ing.com}
\orcid{0009-0004-0901-5388}

\author{Luis Cruz}
\affiliation{%
  \institution{Delft University of Technology}
  \city{Delft}
  \country{The Netherlands}
}
\email{l.cruz@tudelft.nl}
\orcid{0000-0002-1615-355X}

\author{Diomidis Spinellis}
\affiliation{%
  \institution{Delft University of Technology}
  \city{Delft}
  \country{The Netherlands}
}
\email{d.spinellis@tudelft.nl}
\orcid{0000-0003-4231-1897}

\author{Arie van Deursen}
\affiliation{%
  \institution{Delft University of Technology}
  \city{Delft}
  \country{The Netherlands}
}
\email{arie.vandeursen@tudelft.nl}
\orcid{0000-0003-4850-3312}


\begin{abstract}
Effective IT change management is important for businesses that depend on software and services, particularly in highly regulated sectors such as finance, where operational reliability, auditability, and explainability are essential. A significant portion of IT incidents are caused by changes, making it important to identify high-risk changes before deployment. This study presents a predictive incident risk scoring approach at a large international bank. The approach supports engineers during the assessment and planning phases of change deployments by predicting the potential of inducing incidents. To satisfy regulatory constraints, we built the model with auditability and explainability in mind, applying SHAP values to provide feature-level insights and ensure decisions are traceable and transparent. Using a one-year real-world dataset, we compare the existing rule-based process with three machine learning models: HGBC, LightGBM, and XGBoost. LightGBM achieved the best performance, particularly when enriched with aggregated team metrics that capture organisational context. Our results show that data-driven, interpretable models can outperform rule-based approaches while meeting compliance needs, enabling proactive risk mitigation and more reliable IT operations. 
\end{abstract}

\ccsdesc[500]{Software and its engineering~Software reliability}
\ccsdesc[500]{Software and its engineering~Software post-development issues}
\ccsdesc[500]{Computing methodologies~Natural language processing}
\ccsdesc[500]{Computing methodologies~Supervised learning by classification}
\ccsdesc[500]{Computing methodologies~Classification and regression trees}
\ccsdesc[100]{Social and professional topics~Governmental regulations}

\keywords{change management, incident management, reliability, predictive models, classification}

\received{29 September 2025}
\received[revised]{8 December 2025}

\maketitle
\section{Introduction}
At present, major industries such as finance, healthcare, and retail are increasingly reliant on software and related services. However, the reliability and availability of these services are often compromised by \emph{incidents}: unplanned interruptions to a service or reductions in service quality~\cite{iso_standard}. A significant portion of incidents stem from IT changes, which involve additions, modifications, or deletions to existing IT applications~\cite{iso_standard}. As reported by Google~\cite{beyer2016site}, roughly 70\% of outages in its live systems are due to such changes.

To monitor these change-induced incidents, we often refer to the change failure rate, which measures the percentage of deployments causing a failure in production~\cite{ForsgrenHumbleKim18}. It is important to ensure service reliability and availability, because ineffective incident management can lead to customer dissatisfaction, financial losses, and reputational damage. For example, infrastructure failures may cost \$100,000 per hour, while critical application failures can reach \$1 million per hour~\cite{elliot2014devops}.
To mitigate these risks and prevent severe business disruptions, financial institutions in Europe are required to follow rigorous change and incident management processes, ensuring that all changes are carefully assessed and auditable~\cite{eba2022_dora, barta2018increasing}.

Identifying potential incident-inducing changes before deployment can help prevent incidents and ensure service reliability~\cite{ghosh2022fight}. Engineers prefer proactive measures~\cite{kapel2024enhancing}, such as enhancing testing and issue detection early in the development process, a practice known as “Shift Left”. This approach improves service reliability by detecting incident-causing bugs before they reach production~\cite{ghosh2022fight}. Incident prevention entails actively identifying potential failures and forecasting severe outages by using historical data and analytics to predict and mitigate risks~\cite{remil2024aiops}. Predictive incident management is enabled by artificial intelligence for IT Operations (AIOps), which uses machine learning (ML) and big data mining to forecast potential system malfunctions by analysing historical patterns~\cite{remil2024aiops}. 

Ensuring trust, involving humans in the decision-making process, and providing interpretability and explainability of AIOps solutions is essential for gaining confidence from industry practitioners~\cite{remil2024aiops}. Interpretable models are often favoured over black-box models, even when slightly less accurate, as they offer great transparency and support informed decision-making. 

Deploying predictive incident management in a highly regulated financial environment introduces unique challenges. Change and incident management processes in financial institutions are tightly governed by regulatory standards that demand compliance, auditability, and traceability~\cite{eba2022_dora, ai_act_eu2024, barta2018increasing}. These constraints limit the use of black-box models like deep neural networks or large language models~\cite{sapkota2025comprehensive}, which often lack the transparency needed for both regulatory approval and practitioner trust.

In this work, we address these challenges by developing an approach that predicts the probability that an IT change will cause incidents. We focus on boosted tree-based classifiers (HGBC, LightGBM, and XGBoost), which are proven in practice~\cite{nielsen2016tree}, well-suited for tabular IT management data and support post-hoc interpretability through established methods such as SHapley Additive exPlanations (SHAP). Our approach generates an incident prediction score for each planned change, enabling engineers to assess deployment readiness. 
SHAP-based explanations provide feature-level transparency that supports user trust, informed decision-making and meets financial-sector audit requirements. Furthermore, we enrich the dataset with aggregated metrics that reflect team performance, such as change success rates, incident counts, and release outcomes, to evaluate their impact on model performance.

A key design consideration is the trade-off between accuracy and explainability. Regulatory constraints make high-performing but opaque models unsuitable, as change-approval decisions must remain traceable and auditable. The resulting prediction scores produced by the models are intended as risk signals, helping teams prioritise which changes merit closer review.

We evaluate our approach using a one-year dataset from our case company \ifdefined\DOUBLEBLIND ABC (A Banking Company)\footnote{anonymised for submission}\else ING (International Netherlands Group)\fi, a multinational banking and financial services corporation. Building on insights from our prior work~\cite{kapel2024difficulty}, which analysed the characteristics of incident-inducing changes, we tailor our solution to the organisational and regulatory requirements of financial institutions. This ensures the model is not only accurate, but also usable, trustworthy, and compliant for engineers making deployment decisions -- aspects that are essential in our context but have received little attention in previous work.

Our main goal is to equip engineers with a predictive and explainable model that complies with regulatory standards, allowing them to identify high-risk changes during the assessment \& planning phase and take preventive action to ensure reliable deployments. In particular, we seek to improve upon methods based on business rules and human effort, as currently employed at \ing, by training ML models on historical data. We also investigate how team performance metrics can contribute to assessing incident prediction.

We address the following research questions:
\begin{itemize}
\item \textbf{RQ1}: What is, in practice, the performance of a rule-based approach for obtaining the incident prediction score for deployment changes?
\item \textbf{RQ2}: How can we use data-driven ML models for incident prediction scoring, and how does their performance compare to the rule-based approach? 
\item \textbf{RQ3}: What is the effect of including aggregated team metric data on the accuracy of ML-based models for incident prediction scoring?
\end{itemize}

In our study, we demonstrate how data-driven ML methods can be employed within a tightly regulated change management process, providing more effective risk assessment than the rule-based approaches currently used in practice. Among the evaluated models, LightGBM delivers the best performance, particularly when enriched with aggregated team metrics. The model's predictions remain explainable, with influential factors including team metadata, the machine involved, and other product-specific risk indicators. By combining predictive performance with transparency, this approach has the potential to strengthen IT system reliability while reducing the time and resources spent on incident management. To our knowledge, this is the first study to emphasise both feature-level explainability and the integration of aggregated team metrics for assessing change deployment risk in a financial context. 

To sum up, this work contributes  the following.
\begin{enumerate}
    \item A comparative evaluation of a rule-based approach and three ML classifiers (HGBC, LightGBM, and XGBoost) for predicting incident risk of IT changes, using a one-year real-world dataset from a large bank. LightGBM achieves the highest weighted recall and F2-measure. 
    \item A method for achieving enhanced model transparency by applying SHAP values to interpret predictions. This shows that textual change descriptions and team-related metadata are among the most influential features, providing actionable insights that support compliance and informed decision-making.
    \item An analysis of model performance stability over time, demonstrating consistent weighted F2-measure over time.
    \item Evidence that incorporating aggregated team metrics yields modest but meaningful performance improvements, especially in AUC. This underscores the potential of enriching predictive models with other data sources to improve accuracy.
\end{enumerate}


\section{Related Literature} \label{sec:literature} 
Predictive models face several significant challenges, such as the lack of ground truth labels, the necessity for manual effort to obtain high-quality data, highly imbalanced datasets, and complex dependencies among components and services~\cite{remil2024aiops}. Despite this, substantial progress has been made in developing predictive AIOps models, particularly for incident detection and prediction. This paper examines the impact of changes on incidents, categorised into two main areas: 1) pre-change risk analysis and incident prevention~\cite{batta2021system, guven2016understanding}; and 2) post-change identification of failed changes~\cite{zhao2021identifying, zhao2023identifying, li2020gandalf, zhai2020check, yan2023aegis}. Our focus is on pre-change incident identification.

Pre-change incident identification involves predicting whether a change will result in an incident before it is deployed, based on similar historical changes that induced incidents. Changes are a common cause of incidents in live systems, responsible for up to 70\% of outages~\cite{beyer2016site}. Large-scale software companies have many services and resources consisting of numerous components~\cite{chen2020incidental}, which obscure a complete view of the entire system and its relationships. These dependencies make it difficult to predict when a change will induce an incident, especially since seemingly successful changes often lead to incidents~\cite{guven2016understanding, zhao2021identifying}.

Previous work on this topic is limited due to the challenges in systematically collecting data on changes inducing incidents~\cite{guven2016towards}. More focus and time pressure are placed on resolving incidents than on the procedural guidelines that require proper administration~\cite{kapel2024enhancing}. \ifdefined\DOUBLEBLIND Work conducted at ING bank\else Previous work at \ing\fi~\cite{kapel2024difficulty} indicates the complexity of determining links between changes and incidents, emphasising the necessity of handling data imbalance issues when utilising ML on this particular data. 

On the research side, Batta et al. proposed a risk management system based on supporting evidence from past bad changes~\cite{batta2021system}. They trained multiple classification models, including logistic regression, random forest, passive aggressive, support vector machine, and LSTMs (Long Short-Term Memory models), to distinguish problematic changes from successful ones, evaluated on IBM data. Similarly, Güven and Murthy analysed IBM data to identify change-incident linkages and conducted predictive analytics to reduce change-related incidents~\cite{guven2016understanding}. They applied various machine learning algorithms to the properties of a change, finding that classification and regression trees achieved the highest recall.

Ahmed et al.~\cite{ahmed2023empirical} demonstrated that XGBoost is highly effective for predicting IT incident severity, often achieving performance comparable to advanced deep learning models such as BERT, RoBERTa, or ERNIE~2.0. However, their study focused exclusively on major incidents and analysed incidents after they had occurred. In contrast, our work targets high-priority incidents, including both priority~1 and priority~2 events, and predicts the risk of a change before an incident occurs. Furthermore, we extend the feature space with aggregated team metrics, providing richer context. This enables a predictive and interpretable model that leverages the proven effectiveness of tree-based classifier models while aligning with regulatory requirements for pre-deployment risk assessment in financial IT operations.

Looking at industry solutions, ServiceNow provides a cloud-based platform that aids in digitising any business process, with offerings that include IT service and IT operations management~\cite{snow_what}. One of its features provides data-driven insight into the risk during the deployment of a change. However, its purpose is to speed up the change approval process by automatically evaluating and applying approval decisions based on business requirements~\cite{snow_morrison_2021}. While ServiceNow looks at historical change data for insights into team performance~\cite{snow_change_success}, we analyse historical change, incident, and aggregated metric data for insights into team performance. However, due to the proprietary nature of ServiceNow's risk assessment feature and the lack of publicly available technical details, we were unable to directly compare our model with theirs. 

A key distinction of our approach is its focus on explainability. It offers engineers transparency by providing insights into which features influence the scoring and informing users about what determines their deployment risk prediction. Since all approval decisions at \ing require a human-in-the-loop, this additional insight into the scoring aids the approval step in the change deployment process, reinforcing both operational effectiveness and regulatory compliance.

\section{Background} \label{sec:background}
This section provides an overview of the case company, its change management process, and the risk-related factors influencing it.

\subsection{Case Company}
Our research examines \ing, a global financial services company offering financial products and services to millions of customers. With over 15,000 engineers deploying thousands of monthly changes, \ing has evolved from a traditional bank into a digital platform offering online and mobile banking services. The increasing role of ICT in finance necessitates its integration into daily operations~\cite{eba2022_dora}. Consequently, \ing can be categorised as a financial software-defined business~\cite{kapel2024enhancing}. 

Banks play a systemic role in the global economy and are subject to regulatory oversight to ensure they comply with risk guidelines and policies. Regulatory agencies, particularly the European Banking Authority (EBA), significantly influence the management of \ing's processes, given its European base~\cite{kapel2024difficulty}. The EBA aims to ensure effective and consistent prudential regulation and supervision across Europe's banking sector~\cite{eba_aboutus}. Regulatory milestones such as the revised Payment Services Directive introduced in 2017~\cite{eba_2021_major} and the 2019 Guidelines on ICT and security risk management~\cite{eba2019} have significantly influenced IT operations in the financial sector. 

In 2023, the European Parliament adopted the Digital Operational Resilience Act (DORA)~\cite{eba2022_dora}, establishing a regulatory framework for digital operational resilience, ensuring technological safety, proper functioning, and quick recovery from ICT breaches and incidents, which took effect in 2025. DORA aims to enable the effective and smooth provision of financial services while preserving consumer and market trust and confidence. It specifically addresses ICT risks through rules on ICT risk-management capabilities, incident reporting, operational resilience testing, and monitoring of ICT third-party risks.  

\subsection{Change Management Process}
Change management is the process responsible for handling change requests and managing associated risks~\cite{ibm_itil}. At \ing, this process follows the Information Technology Infrastructure Library (ITIL)~\cite{ibm_itil} guidelines, combined with agile principles~\cite{beck2001agile}, enabling multi-disciplinary teams to be responsible for entire processes and value chains, end-to-end~\cite{kapel2024enhancing}. The focus is on IT changes affecting services such as hardware, networks, middleware, and software. Each change is assigned to a single team, affects at least one configuration item (CI), and has a specific implementation moment. The configuration management database (CMDB) tracks all IT assets, processes, and changes to their attributes and relationships~\cite{ibm_itil}.

This process is closely tied to the incident management process~\cite{ibm_itil}, which manages the life cycle of all incidents. When a change is required to resolve an incident, it must be logged and processed through the change management process. Conversely, incident management is responsible for the detection and resolution of incidents that may arise from unsuccessful changes. 

The change management process involves five stages:
\textit{1) Logging}: Registering IT change activities as change tickets in the service management tool, with detailed descriptions to aid approvals and information sharing. 
\textit{2) Assessment \& Planning}: Evaluating the risk of a change, focusing on the probability of failure and potential damage if the change fails. This stage involves noting dependencies with other teams and changes, and automating deployment impact analysis and risk calculation. 
\textit{3) Approval}: Approval groups assess the readiness of a change for deployment. The product owner must approve any change released to production, accepting the delivered quality and risks. Unapproved changes are cancelled.
\textit{4) Coordinate Implementation}: Deploying the change within the agreed time window. 
\textit{5) Evaluation \& Closure}: Evaluating the change post-implementation to determine if it functions as expected. If not, the rollback or remediation plan should be executed. A closure code is registered to indicate if the change was successful, successful with problems, failed, or cancelled. 

Our study focuses on the \textit{Assessment \& Planning} and \textit{Approval} stages by introducing incident prediction scores to help engineers determine whether a planned change is ready for deployment. Accurate prediction at this stage can reduce the likelihood of incident-prone changes reaching production.

\subsection{Risk-related Influences to the Process}
As a financial services provider, \ing must ensure reliability and business continuity~\cite{kapel2024difficulty}. Failed changes can lead to service disruptions, which negatively impact the business. Therefore, robust change deployment management is needed to ensure compliance with governance, legal, contractual, and regulatory requirements. 

Under article 9 of DORA~\cite{eba2022_dora}, financial entities need to continuously monitor and control the security and functioning of ICT systems and tools~\cite{eba2022_dora}. They must minimise the impact of ICT risks by designing, procuring, and implementing security policies, procedures, protocols, and tools to ensure ICT system resilience, continuity, and availability. Specifically, for production, documented policies, procedures, and controls based on a risk assessment approach are required~\cite{eba2019}. These should be integral to the financial entity's overall change management process to ensure all changes are recorded, tested, assessed, approved, implemented, and verified in a controlled manner~\cite{eba2022_dora}. The risk assessment should consider potential impacts on the continuity and quality of financial services~\cite{eba2019}. Furthermore, post-change follow-ups should be conducted to verify the successful implementation without unexpected impacts or the need for remediation~\cite{eba2022_dora}. Our research supports the risk assessment approach by predicting if a change is predicted to cause an incident, thus aiding the decision on whether the change should be deployed in its current state. 

Additionally, the EU's AI Act~\cite{ai_act_eu2024}, introduced in early 2024, establishes a legal framework for the development, market placement, putting into service, and use of AI systems within the EU. The goal is to promote the adoption of human-centric and trustworthy AI. To align with this, we emphasise human-in-the-loop decision-making and explainability in our predictions to build user trust and improve actionability.

\section{Data}\label{sec:data}
We analyse 175k closed change tickets deployed in the company's production environment over one year (November 2022 to October 2023), linked to incident tickets from the same period filtered by priority levels 1 and 2 (out of 5). Priority~1 incidents are characterised by high urgency, requiring to be solved as soon as possible, and high impact, indicating critical effects company-wide or across multiple business domains. Major incidents are identified as special cases of priority 1 that have a major impact on critical business processes or services~\cite{ibm_itil}. Priority~2 incidents have a medium impact on a single business domain and medium urgency. A change is considered to have triggered an incident if its identifier appears in the incident's \textit{Caused by Change} field or is mentioned in its \textit{Solution} field as the cause. This results in about 2.4\% (around 4k) of changes being identified as incident-inducing. 

We further enrich the data with aggregated team metrics, specifically for the team assigned to implement each change. These metrics are derived from incidents, changes, and releases associated with that team. Release data is only available for 23.1\% of changes, providing an overview of all production releases executed through pipelines and their compliance status. For each release, there is information on team metadata, start and end times, release outcomes (success, partial success, or failure), and control columns such as product owner approvals, peer code reviews, and related changes. Aggregated team metrics are linked to change tickets via the \textit{IT Product}, which represents the deliverable owned by a single engineering team.  

To protect confidentiality, we report only aggregated dataset characteristics rather than detailed feature statistics.

\section{Methods}\label{sec:methods}
To enhance change deployment reliability, we propose generating a predictive incident prediction score during the planning phase. We start by reviewing the company's current risk assessment approach, then introduce and evaluate three classification-based ML models. We identify the best-performing model and further test its accuracy by incorporating additional features.

\subsection{Baseline: Rule-Based Scoring}\label{sec:baseline}
The company's change deployment risk assessment is supported by an automated calculation within its IT Service Management (ITSM) tool. This calculation is configured by the company and is based on data available in the tool and the CMDB. The resulting score provides an indication of the probability of failure for the planned change and the potential impact. The calculation result depends on the quality of the information in the change tickets and the CMDB, making it reliant on human effort and expertise to ensure detailed and accurate tickets. 

The ITSM tool uses specific criteria, or business rules, to derive a risk category, considering factors of the probability of failure (e.g., the scope of impacted IT services, deployment complexity, and incident history) and of potential damage (e.g., Confidentiality, Integrity, and Availability ratings, SOx criticality~\cite{sox02}, and recoverability). The output is a score from 0 to 100, mapped to risk levels: low (0-33), medium (34-59), and high (60-100).

This score is generated during the \textit{Assessment \& Planning} stage and informs the \textit{Approval} stage, where the approval group uses it to evaluate deployment readiness~\cite{kapel2024enhancing}. This supports the four-eye principle, a widely adopted practice in financial institutions and other safety-critical domains, which requires at least two independent reviewers to approve each change before deployment. Due to confidentiality constraints, we cannot provide a deeper description of the exact business rules, weightings, or configuration logic used in the current baseline.

\subsection{Scoring using ML models} \label{sec:method_ml}
To improve the current rule-based assessment, we apply data-driven ML methods to historical change and incident data from \ing's ITSM system. Unlike the baseline, which depends on manually entered data and expert judgment, ML models learn patterns from past deployments that could negatively impact stability and cause incidents, reducing subjectivity and improving efficiency. This approach helps engineers and change managers maintain high service quality and fosters continuous improvement in a fast-moving environment. Also, it helps them by ensuring that significant changes receive closer scrutiny, particularly in areas such as the execution plan, testing, and rollback, while allowing lower-risk changes to proceed with less oversight. This allows for streamlining approvals by IT leads.  This is particularly beneficial for Deployment Change Advisory Boards (DCABs), also known as Change Advisory Boards (CABs). 

In addition to predictive scoring, we offer explainability of the models by employing SHAP, a unified framework for interpreting predictions~\cite{lundberg2017unified}. SHAP values enable us to provide users with insights into a single prediction of a planned change. They show how each feature contributes positively or negatively to a given prediction, and identify the most important features that drive the ML models' predictions. SHAP was chosen because it is particularly well-suited for boosted tree models~\cite{lundberg2018consistent}, provides individualised explanations for each prediction, and is widely used in industry and research~\cite{mosca2022shap}. We report the top 15 most important features influencing the predictions of the best-performing ML model. This approach aligns with other ongoing research at \ing, where SHAP is similarly used to provide explanations of an ML model's decision to end users~\cite{roelofs2024finding}. 

By explaining predictions, we increase user understanding and acceptance~\cite{adadi2018peeking, remil2024aiops}. From a regulatory perspective, explainability also ensures that model-driven decisions remain auditable and accountable, supporting financial sector requirements for transparency, traceability, and human oversight. 

\subsubsection{Feature Preparation}
Before modeling, we perform extensive data processing and feature engineering to ensure the quality and relevance of our data.

For processing the incident data, we begin by removing duplicates and filtering out irrelevant incidents based on the company's closure codes, such as `Invalid event' or `Withdrawn by Customer'. Major incidents are a sub-category of Priority~1 incidents, but we introduce them separately as Priority~0. To enrich the data, we check if a change identifier is mentioned in the incident ticket as a cause and add this identifier to the \textit{Caused by Change} field if not already present. 

For processing the change data, we focus on closed changes. 
In preparing the features, we start by removing stopwords from the \textit{Short Description} and \textit{Description} columns of the changes. We then generate Natural Language Processing (NLP) features using scikit-learn’s~\cite{scikit-learn} CountVectorizer, which are fed into a TruncatedSVD for latent semantic analysis. Date features are generated from the change start timestamp, including the starting hour, day of the week, quarter, month, day of the year, day of the month, week of the year, and whether it is a weekend.

We merge both datasets, incorporating all categorical, numerical, date, and NLP features. For categorical features, we include metadata of a change, such as attributes of the CI, security ratings (confidentiality, integrity, and availability), change management information (change category, change state, CAB approval group), incident and support management information (e.g., support offering or assignment group), compliance and criticality information, and deployment and architecture information (automated deployment, fallback options, and redundant architecture). The data is merged with incident data to determine which changes caused incidents by linking the \textit{Caused by Change} incident fields with change data. Connections where the change occurred after the incident are removed to ensure only causal links are retained.

As described in Section~\ref{sec:data}, we incorporate aggregated team metrics as features, derived from incident, change, and release data. These features consist of median weekly and monthly aggregates of team-level performance metrics, computed using the \textit{IT Product} as a proxy for team identity. 
To capture team performance, we compute aggregate metrics such as the number of changes, percentage of successful changes, number of changes causing incidents, number of high-priority incidents, percentage of successful releases, and number of releases. However, since the \textit{IT Product} field currently has only 50\% coverage (with ongoing efforts to improve this), we exclude tickets lacking this information in our analysis for RQ3. 

\subsubsection{Approach}
The \textit{Caused by Change} field labels whether a given change results in an incident. A change is labelled as causing an incident if it has at least one link to a high-priority incident (i.e., major, priority 1, priority 2), independent of how many incidents are linked. For each change, we also record the highest incident priority, which is later used for sample weighting. This labelling process results in a binary classification, indicating that a change has either induced at least one high-priority incident (1) or not (0).

Based on these labels and the prepared features, we train multiple classification ML models on a training set. We split the data temporally, using the first eight months for training (120K changes), then two months for validation (26K changes), and the last two months as the test set (30K changes). We implemented three gradient boosting algorithms in Python: HistGradientBoostingClassifier from scikit-learn~\cite{scikit-learn} (referred to as HGBC), LightGBM~\cite{ke2017lightgbm}, and XGBoost~\cite{chen2016xgboost}. We focused on classification models that can be explained post-hoc to support explainability and auditability, which is a critical requirement in our context. 

Simpler models were unable to handle the Not a Numbers (NANs) inherent in the data,  which provide meaningful information regarding unfilled fields. To preserve this information, we opted for models capable of handling such values, unlike other models like linear regression, support vector machine or random forest classifiers.

Through iterative training, the model learns which features are useful indicators for predicting the labelled outcome. The target variable is an incident prediction score between 0 and 100, indicating the probability that the planned change will trigger an incident. 

\section{Evaluation}\label{sec:evaluation}
Each model serves as a classifier that produces a numerical score between 0 and 100, indicating the probability that a change will trigger a high-priority incident. To convert these into binary predictions, we identify an optimal threshold using a temporal split: eight months for training and two months for validation. The threshold that maximizes the weighted F2-measure (defined below) on the validation set is selected and applied to a held-out two-month test set. 

To simulate real-world deployment, we evaluate the best-performing model using a sliding window setup. In practice, new changes and incident data are generated continuously, and risk assessments must remain aligned with the latest information. Therefore, the model is retrained daily, incorporating the most recent data, and produces predictions on a weekly cycle to support the organisation’s change planning and approval processes. From each prediction window, we derived confusion matrices and computed evaluation metrics to monitor performance and stability over time. 

Precision, defined as $\frac{TP}{TP + FP}$, captures how many positive predictions are correct.
Given the significant class imbalance due to the rarity of high-priority incidents triggered by a change, we report weighted variants of recall and F-measure to more fairly assess model performance. Weighted recall or wR (as defined in Equation~\ref{eq:recall}) averages recall per class, weighted by the number of true instances in each class:

\begin{equation} \label{eq:recall}
\text{wR} = \sum_{c \in C} \frac{n_c}{N} \cdot \frac{TP_c}{TP_c + FN_c}
\end{equation}

Here, $C$ is the set of classes (high-priority incident/ no high-priority incident), $n_c$ the number of true instances of class $c$, and $N$ the total number of instances.

To reflect the importance of minimising false negatives so that we do not miss incident-inducing changes when they are being planned, we compute the weighted F2-measure or w$F_2$ (as defined in Equation~\ref{eq:f1}). Greater emphasis is placed on recall using $\beta$=2, which is inspired by prior work~\cite{batta2021system}. 

\begin{equation} \label{eq:f1}
\text{w}F_\beta = \sum_{c \in C} \frac{n_c}{n_0 + n_1} \cdot \left( \frac{(1 + \beta^2) \cdot P_c \cdot R_c}{\beta^2 \cdot P_c + R_c} \right)
\end{equation}

Where $P_c$ and $R_c$ are the class-specific precision and recall scores.

Our evaluation aligns with the \textit{Assessment \& Planning } stage of the change management process, where the objective is to proactively flag changes that trigger incidents before deployment. A high recall is crucial to ensure that most incident-inducing changes are identified, even at the cost of some false positives. These predictions are accompanied by model explanations to assist change managers and engineers in making informed decisions on whether to proceed and if additional actions are needed. By reviewing the prediction explanations, we ask for more due diligence from the engineer to avoid incidents similar to those in historic changes. Therefore, a high recall (even at moderate precision) supports risk-aware decision-making.

Finally, to account for the practical significance of varying incident severities, we introduce priority-based weighting in our evaluation. Changes that lead to major or priority 1 incidents receive a weight of 5, those resulting in priority 2 incidents receive a weight of 3, and non-incident changes are weighted as 1. This ensures that the evaluation process gives proportionally more importance to correctly identifying severe incidents.

\section{Results} \label{sec:results}
This section compares the effectiveness of the company's existing rule-based risk assessment against the ML methods. We first evaluate the baseline to establish a performance benchmark, then assess how the ML models improve predictive accuracy. We also examine the stability of the best performing model its performance over time, and examine the impact of incorporating aggregated team metric data.  

\subsection{RQ1 - Performance of Rule-Based Approach}
At \ing, changes with a risk score of 60 or higher are considered a high-risk level during the \textit{Assessment \& Planning} stage of the change deployment management process. In this paper, we define a high-risk level to indicate that the case company considers a given change as potentially resulting in an incident. Using this threshold of 60 on the two-month test set, the baseline performance for all metrics is presented in Table~\ref{table:combined_perf}.


\begin{table}[htbp]
\centering
\caption{Performance of Baseline, ML Models, and LightGBM with and without Additional Features}
\label{table:combined_perf}
\begin{tabular}{lccccc}
\toprule
& \multicolumn{3}{c}{} 
& \multicolumn{2}{c}{LightGBM} \\
 \cmidrule(lr){5-6}
Metric 
& Base 
& HGBC 
& XGB 
& Without 
& With \\ 
\midrule
Threshold        & 60 & 99  & 98 & 17 & 29 \\
Precision        & 0.04 & 0.02  & \textbf{0.05} & 0.02 & 0.04 \\
wR  & 0.56 & 0.86 & 0.89 & 0.91 & \textbf{0.93} \\
w$F_2$      & 0.88 & 0.88 & 0.90 & 0.92 & \textbf{0.93} \\
AUC              & 0.55 & 0.61 & \textbf{0.71} & 0.67 & 0.60 \\
\bottomrule
\end{tabular}
\end{table}
 
The baseline approach achieves very low precision, indicating that most changes flagged as high risk do not actually result in incidents. This is largely due to the highly imbalanced nature of the dataset, where incident-inducing changes are rare, making it difficult for static rules to distinguish incident-inducing changes from safe ones. However, the wR and w$F_2$ are better, since the recall identifies 56\% of the relevant cases within the test set, and the w$F_2$ shows the weighted precision and recall, giving a good indication for our business purposes. When examining the area under the receiver operating curve (AUC), the score of 0.56 is close to 0.5, suggesting that the model has almost no discriminating ability~\cite{hoo2017roc} and thus can be interpreted as failing in classifying this task~\cite{nahm2022receiver}. 

\finding{%
The baseline risk assessment fails to effectively identify high-risk changes, as evidenced by low precision, a moderate wR of 56\%, and an AUC of 0.56, indicating poor discriminative performance.}

\subsection{RQ2 - Comparison of ML Models vs. Rule-Based Approach}\label{sec:RQ2}
\subsubsection{Performance}
We evaluated three ML models, HGBC, LightGBM, and XGboost, against the rule-based baseline, optimising each model's threshold on a validation set and assessing final performance on a two-month test set (see Table~\ref{table:combined_perf}). For each performance metric, the best-performing value is highlighted in bold.  


The baseline achieves a precision of 0.04, but is outperformed by XGBoost, which achieves the highest precision (0.05). Both HGBC and LightGBM record lower precision values of 0.02. However, our primary objective is to maximise correct identifications, acting more as a guard dog rather than prioritising the minimisation of false positives. This approach favours recall over precision, making the small loss in precision acceptable. 

In terms of wR, all ML models substantially outperform the baseline. LightGBM achieves the highest value (0.91), followed by XGBoost (0.89) and HGBC (0.86), whereas the baseline reaches 0.56. This performance gap is also apparent in the w$F_2$, which weighs recall more heavily. LightGBM again leads with a w$F_2$ of 0.92, followed by XGBoost(0.90), and HGBC (0.88), with the baseline also scoring 0.88. Despite matching HGBC on w$F_2$ numerically, this value of the baseline is less meaningful given its substantially lower recall, meaning it identifies fewer relevant cases overall. 

The AUC scores provide further insight into each model its overall discriminative ability. XGBoost achieves the highest AUC (0.71), followed by LightGBM (0.67) and HGBC (0.61), while the baseline records the lowest score (0.55). As an AUC score falling between 0.5 and 0.6 means it fails in classifying this task~\cite{nahm2022receiver} and has near-random performance, the ML models demonstrate moderate improvements in classification ability over the baseline. 

There is also considerable variation in decision thresholds across models. HGBC and XGBoost both use very high thresholds (99 and 98, respectively), indicating a low decision boundary likely aimed at maximising recall. In contrast, LightGBM uses a much lower threshold (17), suggesting that its scoring scale differs and may necessitate model-specific calibration. 

Overall, LightGBM offers the best performance in terms of wR and w$F_2$, aligning well with our goal of maximising correct identifications. Despite relatively modest AUC values, all ML models outperform the rule-based baseline and are better suited for high-recall tasks. 

\subsubsection{Important Features}
\begin{figure}[htbp]
    \centerline{\includegraphics[width=0.8\columnwidth]{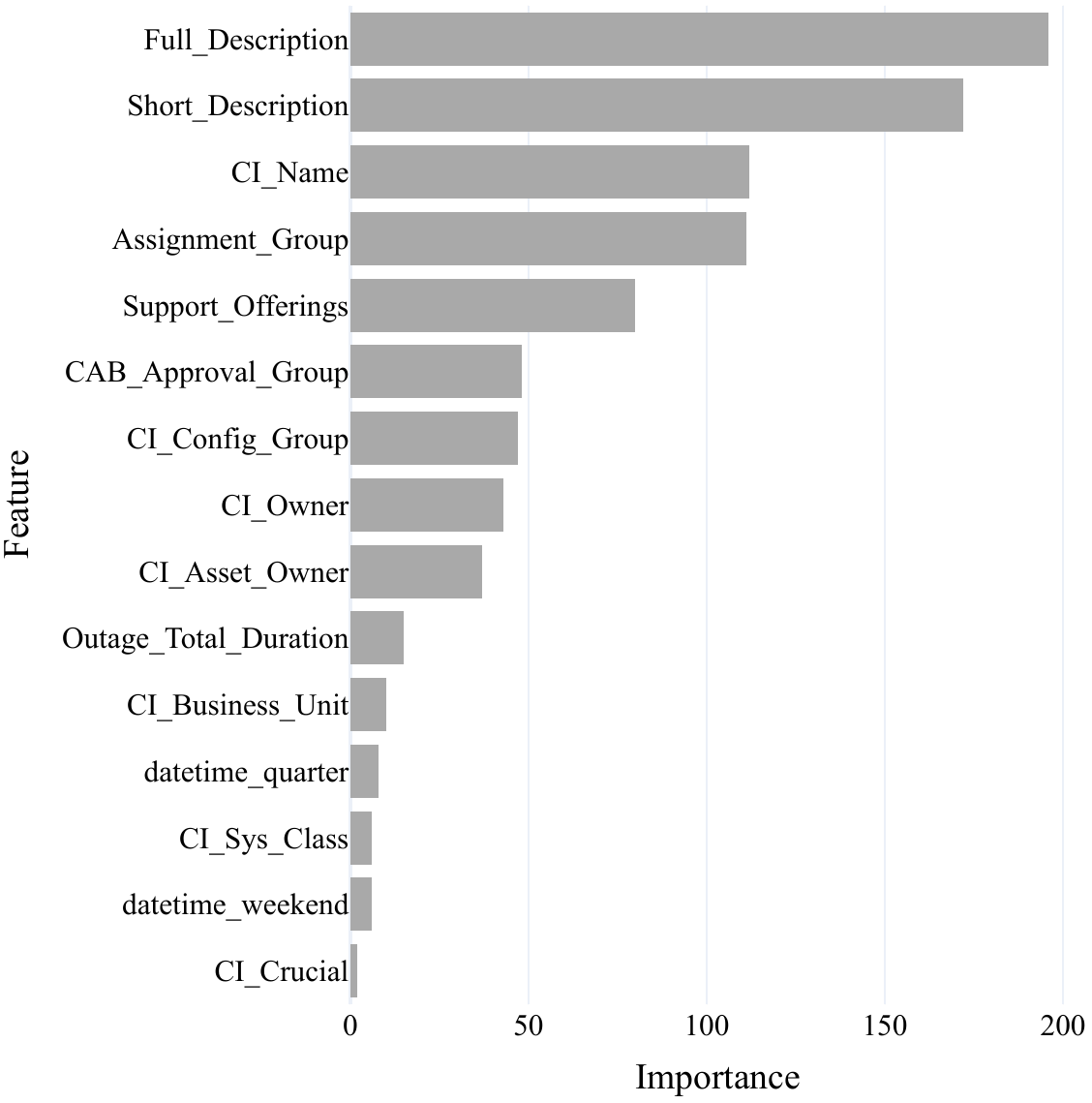}}
    \caption{Feature Importance of LightGBM.}
    
    \Description{Horizontal bar chart showing the top 15 features ranked by SHAP values for the LightGBM model. Higher values indicate a greater importance of a feature to the model’s predictions across the evaluation set. The plot highlights the relative influence of textual, configuration-related, and temporal features in identifying incident-inducing changes (RQ2).}
    \label{fig:feature_importance}
\end{figure}

Next, we examine the best-performing model, LightGBM. Using SHAP, we analyse the top 15 most important features for prediction outcomes, as shown in Fig.~\ref{fig:feature_importance}. Description values have been concatenated due to the abundance of NLP features. Interestingly, the textual features \textit{Full Description} and \textit{Short Description} have the most influence, highlighting that the content of the change significantly impacts the outcome. This is followed by the \textit{CI Name}, which represents the configuration item that is being changed. Additionally, the \textit{Assignment Group}, which refers to the team assigned to work on the change, and the \textit{Support Offerings}, which are the team supporting the CI, are also highly influential. Other features, such as the change approval board group, the CI configuration group, and the owner of the CI, are less important but still influence the predictions. 

\subsubsection{Stability Over Time}
\begin{figure}[htbp]
\centering
\begin{subfigure}[b]{\columnwidth}
    \centering
   \includegraphics[width=0.85\linewidth]{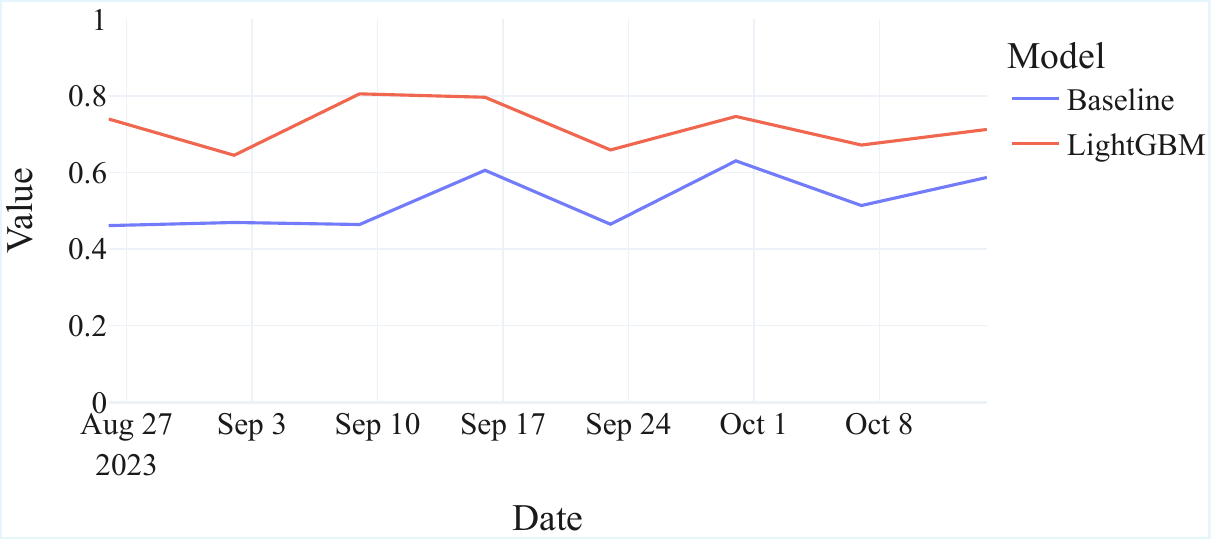}
   \caption{wR}
   \Description{Line chart showing weighted recall over time for the baseline and LightGBM models using a sliding window evaluation. Each point represents performance on a specific window, illustrating how weighted recall varies as the underlying change distribution evolves.}
   \label{fig:sim_recall} 
\end{subfigure}

\begin{subfigure}[b]{\columnwidth}
    \centering
   \includegraphics[width=0.85\linewidth]{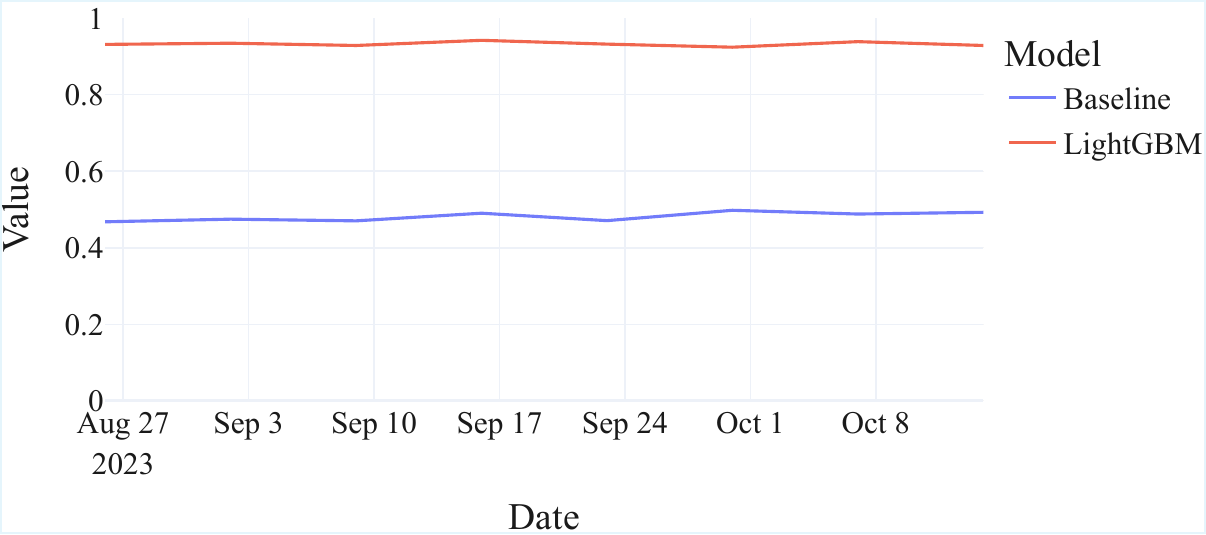}
   \caption{w$F_2$}
   \Description{Line chart showing weighted $F_2$-measure over time for the baseline and LightGBM models using a sliding window evaluation. The plot highlights the temporal stability of model performance, with LightGBM consistently achieving higher scores than the baseline across evaluation windows.}
   \label{fig:sim_f2}
\end{subfigure}

\caption{Sliding Window of Baseline vs. LightGBM Performance.}
\label{fig:sliding_window}
\end{figure}

To further assess the stability of LightGBM, we used a sliding window evaluation to simulate its performance over time, comparing it with the rule-based baseline method (see Fig.~\ref{fig:sliding_window}). This evaluation shows that LightGBM consistently achieves higher wR values compared to the baseline over time, indicating a stronger ability to correctly identify positive cases. Both wR values show some fluctuations, with a peak around early September for LightGBM. This inconsistency is likely due to a period of minimal changes at the company, during which modifications were infrequent and strictly supervised.

In contrast, the w$F_2$ for both models remains relatively stable over time. LightGBM again outperforms the baseline, highlighting its ability to maintain a good balance between precision and recall, which is important for our goal since we want to maximise the identification of correct cases. These results suggest that LightGBM not only delivers performance improvements over the baseline but also maintains its performance in a production-like environment.

\finding{%
Among the evaluated models, LightGBM performs best in wR and w$F_2$, making it the most suitable for high-risk change detection. Key features are primarily textual and team-related. Despite some fluctuations in wR during periods of low change activity, LightGBM shows stable w$F_2$ performance over time.
}

\subsection{RQ3 - Inclusion of Aggregated Team Metric Data}
\subsubsection{Performance}
The best-performing model, LightGBM, is used to evaluate whether the inclusion of additional features from metric data aggregated by \textit{IT Product} improves its performance. The performance is compared with and without these additional features (See Table~\ref{table:combined_perf}).


The addition of team aggregated features results in a higher classification threshold and a noticeable improvement in both precision and wR, increasing from 0.02 to 0.04 and from 0.91 to 0.93, respectively. The  w$F_2$ also improves slightly from 0.92 to 0.93, reflecting that the model is better at balancing between precision and recall. However, the AUC score decreases slightly from 0.67 to 0.60, suggesting that the model's performance based on the optimised threshold improves, but its overall scoring consistency across all instances slightly worsens. Despite this drop in AUC, the gain in precision and recall indicates a more confident identification of positive instances. Integrating data from other sources, with an emphasis on ease of merging, proves to be a worthwhile approach, given the coverage issues discussed in Section~\ref{sec:methods}). 

\subsubsection{Important Features}
\begin{figure}[htbp]
    \centerline{\includegraphics[width=0.8\columnwidth]{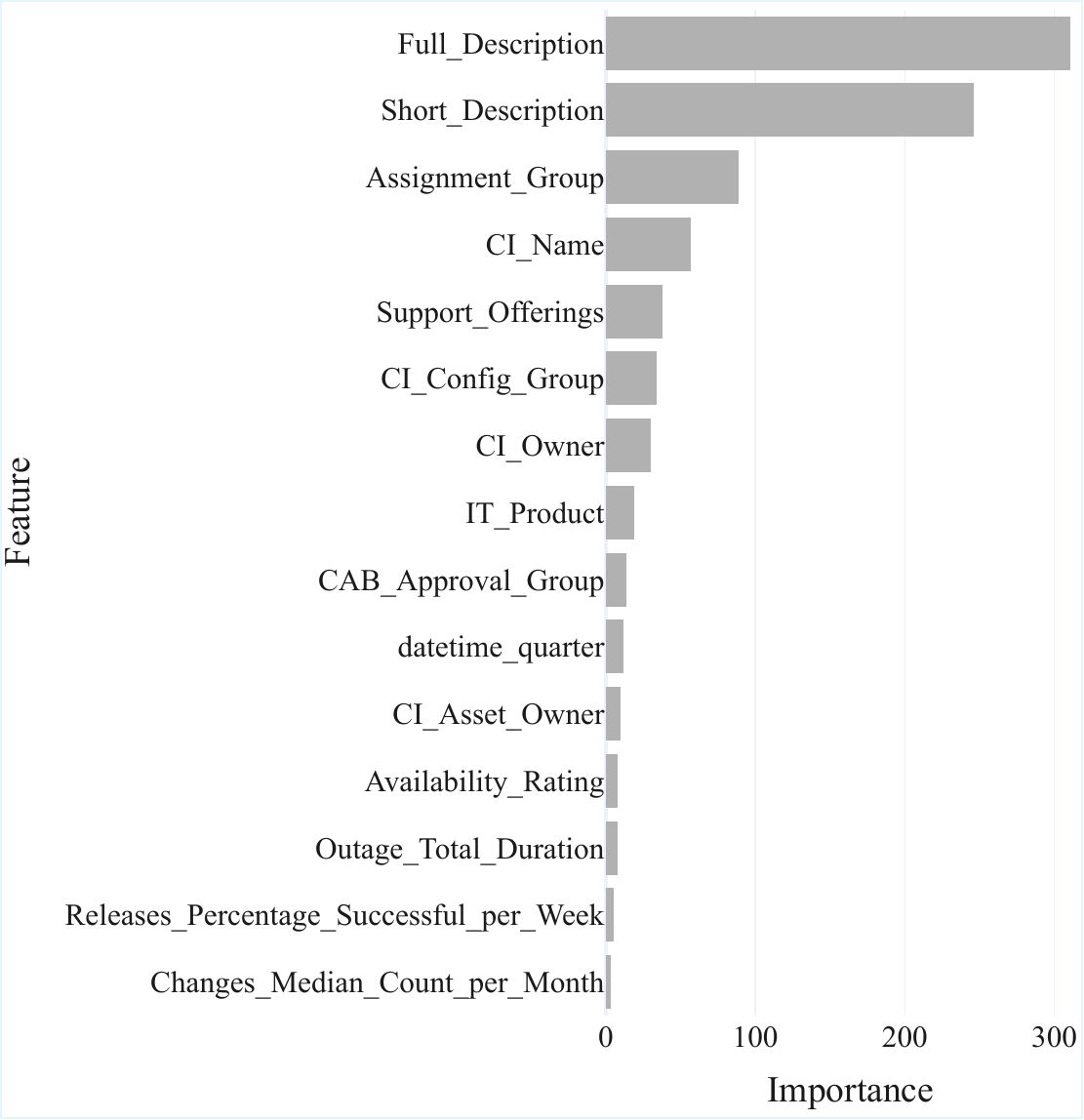}}
    \caption{Feature Importance of LightGBM with Additional Features.}
    
    \Description{Horizontal bar chart showing the top 15 features ranked by SHAP values for the LightGBM model extended with additional features. Higher values indicate a greater importance of a feature to the model’s predictions across the evaluation set. The figure illustrates how textual, organisational, configuration-related, and operational metrics contribute to identifying incident-inducing changes (RQ3).}
    \label{fig:feature_importance_add_features}
\end{figure}

Upon closer inspection of the top 15 SHAP ranking of the model with the added features (see Fig.~\ref{fig:feature_importance_add_features}), we observe the inclusion of these new features. Notably, the \textit{IT Product} ranks 8th, while \textit{Release Percentage Successful per Week} and \textit{Changes Median Count Per Month} are also included, appearing near the bottom of the ranking. 

\subsubsection{Examples}
\begin{figure}[htbp]
\centering
\begin{subfigure}[b]{\columnwidth}
    \centering
   \includegraphics[width=0.9\linewidth]{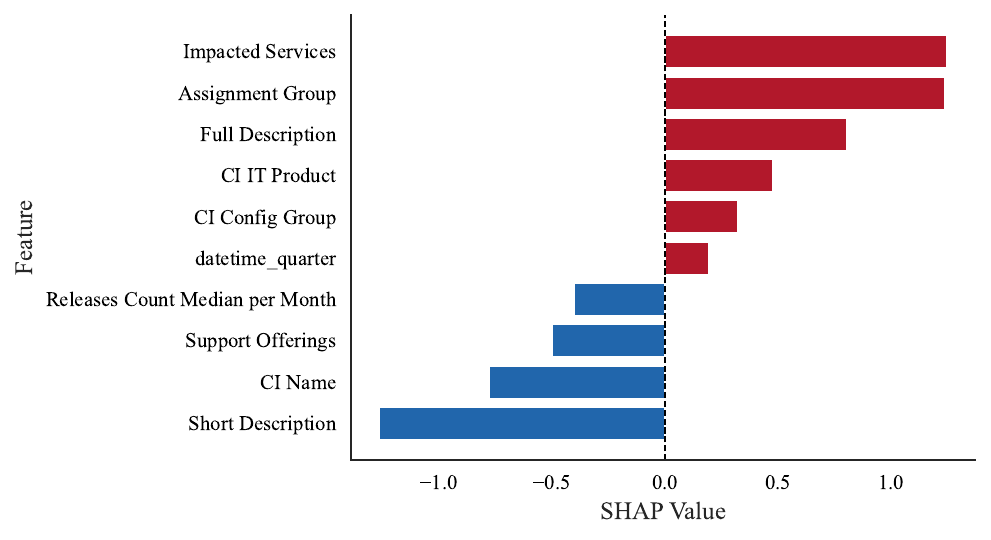}
   \caption{High Scoring Change}
   \Description{Horizontal bar chart showing the top 10 features with the highest SHAP values for a single change instance that receives a high predicted risk score. Bars to the right of zero indicate features that increase the predicted risk, while bars to the left indicate features that decrease it.}
   \label{fig:example_high} 
\end{subfigure}

\begin{subfigure}[b]{\columnwidth}
    \centering
   \includegraphics[width=0.9\linewidth]{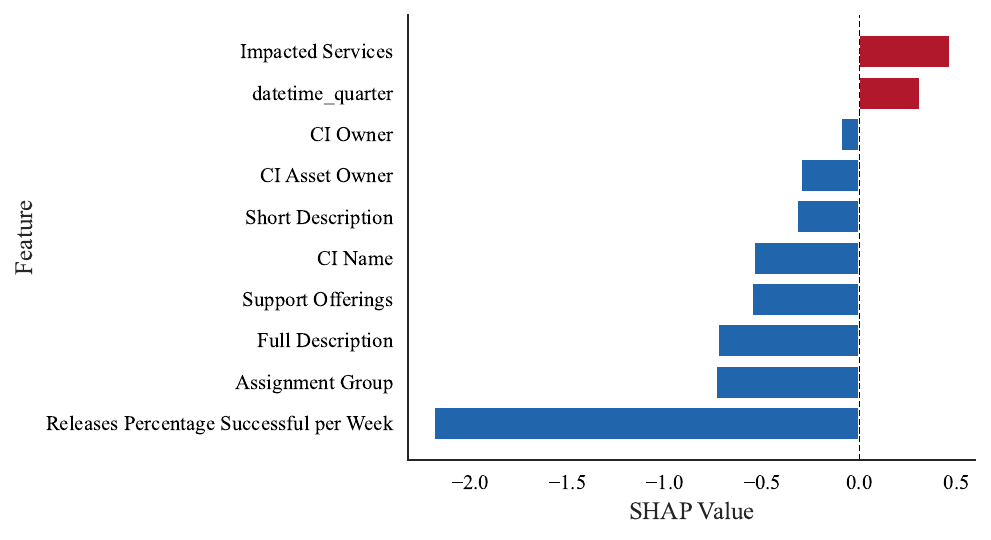}
   \caption{Low Scoring Change}
    \Description{Horizontal bar chart showing the top 10 features with the highest SHAP values for a single change instance that receives a low predicted risk score. Positive SHAP values increase the predicted risk, whereas negative values decrease it.}
   \label{fig:example_low}
\end{subfigure}

\caption{Top 10 SHAP Values for Two Example Changes.}
\label{fig:examples}
\end{figure}

Fig.~\ref{fig:examples} shows two SHAP plots explaining individual predictions. To improve interpretability, we display the ten most impactful features. For multi-component features such as \textit{Full Description} and \textit{Short Description}, we use the maximum signed SHAP value to highlight the most influential element. Features with positive SHAP values (right side)  contribute to a higher predicted risk, while negative values (left side) reduce it, offering users a transparent view into the model's decision-making process by quantifying the contribution of each feature to the output. 

Figure~\ref{fig:example_high} shows a high-scoring change, specifically a monthly infrastructure patch that resulted in a user-facing incident. In this example, \textit{Impacted Services} is the most influential positive feature, pushing the score toward a high-risk classification. Other top contributors, which were also in the top 5 in Figure~\ref{fig:feature_importance_add_features}, include \textit{Assignment Group}, \textit{Full Description}, \textit{Short Description}, \textit{CI Name}, and \textit{Support Offerings}, which appear on both sides of the SHAP scale, suggesting that they have different effects depending on context. Additionally, the team aggregated feature \textit{Releases Count Median per Month} also emerges as an important factor. 

Figure~\ref{fig:example_low} illustrates a low-scoring change involving a routine certificate renewal for network switches, which did not result in any incident. Here, most of the top features contribute negatively, pulling the score downward. Notably, the additional feature \textit{Releases Percentage Successful per Week} has a strong negative SHAP value, indicating its role in de-risking the change. While \textit{Impacted Services} is again present, its influence is limited. The same core features observed in the high-scoring case also appear here, but with negative contributions, which emphasise their dynamic impact depending on the nature of the change.

\finding{%
Adding aggregated team metrics to the LightGBM model yields a modest performance improvement, particularly in precision, wR, and w$F_2$, with new features like IT Product, release success metrics, and change frequency emerging as moderately important predictors.
}

\section{Discussion} \label{sec:discussion}
Both practitioners and researchers can utilise our results for using data-driven ML methods to improve the risk assessment of changes before deployment, while taking into account the corresponding threats to their validity.

\subsection{Implications for Practitioners} \label{sec:dis_pract}
Due to the low amount of incident-inducing changes present in the data, all models inevitably exhibit low precision. This can create operational challenges, most notably alert fatigue~\cite{beyer2016site}. Predictions should therefore not be interpreted as binary signals but as a \textbf{risk-ranking mechanism} that highlights a small subset of changes that need additional attention. To better align with existing approval workflows, practitioners benefit from \textbf{multi-level predictions} (e.g., low/ medium/ high risk), enabling automatic handling of low-risk changes while directing expert attention to the highest-risk ones. To keep such a system usable, a brief \textbf{human-in-the-loop} step can be incorporated in which experts dismiss obvious false positives, apply adaptive thresholds to focus on the highest-risk changes, and provide lightweight feedback (e.g., ``useful/ not useful") to iteratively refine the model. These safeguards enable meaningful decision support without overwhelming engineers or approvers.

Our results show that LightGBM outperforms the rule-based baseline, \textbf{support}ing engineers, change managers, and approvers, especially those with less experience. Whereas experienced practitioners often rely on intuition and tacit knowledge of team behaviours and historical patterns, ML models formalise these insights from historical data. This enables newer users to more accurately assess the risk of their planned changes. 

Model \textbf{explainability} through SHAP values enhances interpretability and practical relevance by linking predictions to understandable features (see Fig.~\ref{fig:examples}). Key features such as \textit{Impacted Services}, \textit{Assignment Group}, and \textit{Releases Percentage Successful per Week} correspond to known risk factors. SHAP plots clarify how each feature influences risk, e.g., a successful release history lowers risk, while involvement of critical services raises it. This makes the model's behaviour more transparent and supports informed decision-making. 

The feature importance results of RQ2 and RQ3 show similar trends. The textual descriptions dominate both, indicating that the content of the change strongly affects the predictions. These are followed by features describing the CI being changed, such as \textit{CI Name} and \textit{CI Config Group}, and the responsible team, like the \textit{Assignment Group} or \textit{Support Offerings}. \textbf{Team-related influences} may stem from factors beyond reliability, such as reporting more faults or handling riskier tasks. Future research should explore these dynamics to prevent bias against specific teams. Interestingly, CAB approval group importance diminishes once aggregated team metrics are added, suggesting that the other team-related features already capture relevant context. 

Contrary to expectations, most rule-based baseline factors were absent from the top 15, except \textit{Availability Rating}. Expected factors such as \textit{Confidentiality},   \textit{Integrity}, and features related to deployment complexity did not significantly influence the predictions. Surprisingly, the features \textit{Datetime Quarter} and \textit{Outage Total Duration}, both time-related, proved influential. This indicates that the specific quarter in which a change occurs influences the deployment risk. One possible explanation is the influence of error budgets~\cite{beyer2016site}, since if a company has already experienced significant user-impacting outages within a given period, it may enforce a freeze on high-risk changes to prevent further service degradation. Such policies directly affect the distribution of risky changes during specific timeframes (see Finding 2), which is then reflected in the model’s predictions. Additionally, the total duration of outages previously experienced by a team significantly impacts the deployment risk of their planned changes.  Added features such as \textit{Release Percentage Successful per Week} and \textit{Changes Median Count per Month} were found to influence the prediction, likely serving as proxies for team maturity.

Features can be grouped into actionable and non-actionable, and this distinction influences how they can be acted upon during the decision if the change is ready to be approved for deployment. \textbf{Actionable features} are those that engineers can directly influence before deploying a change, such as refining the description, improving testing plans, or adjusting rollback strategies, which enables practitioners to address concrete issues highlighted by a high score. In contrast, \textbf{non-actionable features}, such as machine-specific characteristics or team-related metadata, cannot be modified for the current deployment but still provide valuable context. If a particular CI consistently appears in high-scoring changes, this signals structural or historical risk that may warrant additional safeguards, more thorough reviews, phased rollouts, or enhanced monitoring during deployment. Similarly, if certain assignment groups are frequently associated with riskier changes, whether due to the nature of the work or team experience, it may be advisable to involve additional reviewers or support from lower-risk teams to ensure safer deployments.

\textbf{Data quality} significantly influences the outcomes of AIOps methods~\cite{remil2024aiops}. Increasing practical awareness of the importance of linking incidents to changes would yield better labels for model training. Ongoing efforts at \ing have raised awareness about the importance of data quality and robust incident and change management registration to improve reliability. Scaling this across more teams and companies would benefit broader AIOps initiatives. 

Lastly, due to the incomplete \textit{IT Product} field, additional features could only be applied to part of the dataset, limiting the full view of the production environment. Better \textbf{integration of data from other sources}, with an emphasis on ease of merging, proves to be a worthwhile approach.

\subsection{Implications for Researchers}
A major challenge in applying ML to AIOps is the \textbf{data quality in practice}, especially the effort required to link incidents to their causing changes~\cite{remil2024aiops}. Future research should focus on automating both historical and real-time linking, building on prior work~\cite{guven2016towards, batta2021system, kapel2024difficulty}, to enable higher-quality labels and more accurate risk predictions. 
The features identified in this study could serve as a foundation for such research. 

\textbf{Explainability} remains critical for adoption. Researchers should explore how different roles, such as engineers or change managers, prefer to receive and interpret results, and how these can be effectively communicated. Researchers should consider the end-users’ perspective, focusing not only on presenting probabilities but also on how these insights will be utilised in practice.

Finally, the distinction between \textbf{actionable and non-actionable features} highlights the need for further investigation into how models can support practitioners. For example, description fields or deployment timing are actionable, while team metadata is not directly actionable but still provides context for safeguards. Future work should explore how to present both types of features effectively, guiding engineers in reducing deployment risks. Additionally, while SHAP offers valuable interpretability, it has limitations: it explains correlations identified by the model rather than causal relationships~\cite{Lundberg_2021}. Therefore, it cannot reliably indicate which features should be manipulated to achieve a desired outcome. Understanding how to combine interpretability with causal reasoning represents an important direction for research. 

\subsection{Threats to Validity} \label{sec:dis_threats}
The threats and limitations of the study are categorised into construct validity, external validity, and reliability~\cite{yin2009case}.

\textit{1) Construct validity:} We used w$F_2$ to emphasise recall over precision. Other values of $\beta$ can be employed depending on the use case. Adjusting $\beta$ for different scenarios could significantly impact performance outcomes, tailoring the evaluation to different application needs. 
In RQ3, higher precision and recall came with a slight AUC decrease, which is an acceptable trade-off given that AUC can be misleading in imbalanced datasets, as it emphasises overall ranking rather than the correct identification of the minority-class~\cite{ZOU20162}. Low precision is expected because only a small fraction of changes cause incidents, and in this context recall is prioritised, as missing a high-priority incident carries far greater consequences than reviewing additional false positives. 

\textit{2) External validity:} Data-driven methods outperformed the baseline in terms of wR and AUC scores, with LightGBM also exceeding the baseline on the  w$F_2$. These results demonstrate the value of data-driven approaches over approaches that rely predominantly on human judgment. However, they must be interpreted within the context in which the study was conducted. Similar outcomes are expected in other financial software-defined environments, although differences in processes, metadata quality, and taxonomies may influence replication outcomes. Due to confidentiality constraints, detailed feature data cannot be shared; however, our analysis provides insights expected to generalise to comparable settings.

Beyond finance, the methodology can extend to organisations with structured, audited change-management workflows. Highly regulated sectors such as healthcare and government follow similar compliance-driven processes~\cite{mishra2019regulated_change}, and often benefit from explainable, data-driven models that integrate into existing approval processes. Although available features and data completeness vary across domains, the overall workflow (feature extraction, supervised learning, and human-in-the-loop review) remains transferable with domain-specific adaptation of features and thresholds. 

The temporal splitting and class imbalance add further considerations. Operational environments often show seasonal fluctuations (e.g., code freezes), which can lead to distribution shifts between training, validation, and test periods. Because the minority class is very small, even modest month-to-month variation in incident-inducing changes may influence precision and w$F_2$, making temporal evaluation more volatile than random splits. Our long historical training period, two hold-out sets, and imbalance-aware metrics help mitigate these effects, but they remain relevant when extrapolating results.

\textit{3) Reliability} was evaluated via a validation set and a sliding-window evaluation over time. Precision and w$F_2$ remained stable, while variations in wR can be attributed to a strict change phase. Additionally, the challenge of working with an extremely imbalanced dataset, which is common in AI problems~\cite{japkowicz2002class} and observed in other AIOps research on incident management~\cite{remil2024aiops}, was addressed by weighting labels according to the priority of incidents. Alternative strategies, such as class downsampling or upsampling~\cite{japkowicz2002class}, or adjusting the currently used weight distributions, could be explored, but our current approach proved most effective.


\section{Conclusion} \label{sec:conclusion}
This study presents a predictive incident prediction score that can assist engineers during the assessment \& planning phase of IT changes in a demanding production environment. Our results demonstrate that data-driven ML methods outperformed currently employed approaches that rely on business rules and human effort. Among the tested models, LightGBM yielded the best performance, particularly when enriched by aggregated team metrics.

The model's predictions are explainable through SHAP analysis, which identifies the key features influencing outcomes. Notably, metadata related to the team handling the change, the specific machine involved, and other risk indicators inherent to the product were important in shaping the predictions. 

Our findings highlight the value of adopting a data-driven ML approach to improve the reliability of IT systems. In highly regulated environments such as finance, where auditability, traceability, and human oversight are mandatory, this approach provides interpretable risk scores that support compliance. We also highlight the importance of incorporating qualitative metrics, such as team metric features, into change and incident management processes. By enabling more focused scrutiny of changes with a higher likelihood of causing incidents, this approach facilitates proactive risk mitigation and contributes to a more reliable IT environment across the organization.

\begin{acks}
This work was partially supported by \ing through the AI for Fintech Research Lab with Delft University of Technology. 
\end{acks}

\bibliographystyle{ACM-Reference-Format} 
\bibliography{references.bib}

\end{document}